\documentclass[twocolumn,showpacs,preprintnumbers,amsmath,amssymb]{revtex4}
\usepackage{color}
\usepackage{graphicx}
\usepackage{dcolumn}
\usepackage{bm}
\usepackage{hyperref}

\begin{document}

\title{Bidirectional information flow quantum state tomography}

\author{Huikang Huang$^1$}
\author{Haozhen Situ$^{1,}$}
\email{situhaozhen@gmail.com}
\author{Shenggen Zheng$^{2,}$}
\email{zhengshg@pcl.ac.cn}
\affiliation{$^1$College of Mathematics and Informatics, South China Agricultural University, Guangzhou 510642, China \\
$^2$Circuits and Systems Research Center, Peng Cheng Laboratory, Shenzhen 518055, China}%

\date{\today}


\begin{abstract}
The exact reconstruction of many-body quantum systems is one of the major challenges in modern physics, because it is impractical to overcome the exponential complexity problem brought by high-dimensional quantum many-body systems. Recently, machine learning techniques are well used to promote quantum information research and quantum state tomography  has been also developed by neural network generative models. We propose a quantum state tomography method, which is based on Bidirectional Gated Recurrent Unit neural network (BiGRU), to learn and reconstruct both easy quantum states and hard quantum states in this paper.  We are able to use fewer measurement samples in our method to reconstruct these quantum states and  obtain high fidelity.
\end{abstract}

\maketitle

\section{Introduction}
As a fundamental research topic in quantum information processing, quantum state tomography (QST) has always been a concern. As a data-driven problem, QST aims at obtaining as much information as possible of a quantum system and reconstructing the quantum state density matrix through effective quantum state measurement. Actually, accurate QST is impractical \cite{Haffner}, especially cases in high-dimensional quantum many-body systems. One needs exponential complexity to describe a generic quantum many-body system. Even for small-scale quantum systems to be tomographic, it still requires a lot of resources. Therefore, we want to reconstruct the most accurate quantum system through as less measurement samples as possible. In other words, we want to capture the associations between these limited number of quantum state measurement samples and get more information to serve tomography.

After years of investigation, several key technologies has been developed in QST, including, but not limited to, the following methods,  compressed sensing tomography \cite{2010Quantum,QiYin2018}, permutationally invariant tomography \cite{2010Permutationally,2012Permutationally} and tomographic schemes based on tensor networks \cite{2013Scalable}. In recent years, quantum machine learning has  attract a lot  attentions in quantum computing \cite{biamonte2017quantum,situ2020quantum,he2021Variational}.
Meanwhile, machine learning has made great progress in assisting the research of quantum physics problems \cite{carleo2017solving,hartmann2019neural,deng2017quantum,2018wang,leiPhysRevLett,Tang-Shi}. Meanwhile, QST driven by neural network generative models has also received widespread attentions and some relevant research results has appeared. For example, based on probabilistic undirected graph model,  Torlai {\it et al.~}\cite{torlai2018neural} used restricted Boltzmann machine (RBM) QST method to learn amplitude and phase of quantum state and reconstruct  quantum state wavefunction.
Based on the powerful autoregressive model Recurrent Neural Networks (RNN), Carrasquilla  {\it et al.~}\cite{Carrasquilla2019} introduced  RNN-QST, which is able to  use the informationally complete (IC) positive-operator valued measures (POVMs) samples, to reconstruct quantum states with high classical fidelity.  There is also a transformer \cite{2017Attention} QST method based on attention mechanism-based generative network, which reconstructs  mixed state density matrix of a noisy quantum state \cite{Attentionbase2020}. Moreover, there are some other QST methods driven by generative models \cite{CGAN2020,luchnikov2019variational}. All of these methods demonstrate that machine learning techniques can effectively deal with specific quantum states.

In this paper, we propose to use a Bidirectional Recurrent Neural Network (BiRNN) generative model based on contextual semantics to perform QST. By slicing quantum state measurement samples as time series information flow, we are able to  make full use of the contextual semantics of these messages to perform QST by this bidirectional neural network. At the same time, we propose a network training standard to conduct early stopping, which can help one to  find effectively better training model using fewer training samples. These methods enable us to use fewer quantum state measurement samples than RNN\cite{Carrasquilla2019} and AQT\cite{Attentionbase2020} neural network tomography to achieve over 99\% classical fidelity on GHZ state. We test our method in dealing with ``easy quantum states" and ``hard quantum states" that have  different sampling difficulty \cite{rocchetto2018learning}. Finally, we have a brief discussion about why this QST method can effectively process some specific quantum states.

\section{Method}
Our method belongs to unsupervised machine learning. The training samples are produced based on the IC-POVMs mentioned in \cite{Carrasquilla2019}. The IC-POVM operators which we use here are  Pauli-4 operators. Each $N$-qubit quantum state measurement sample is denoted by \textbf{a}. We slice each of them into $T=N$ parts as time series information flow, i.e., $\textbf{a} = [a_1,a_2,...,a_N]$, where $a_i\in\{0,1,2,3\}$. The probability distribution corresponding to the Pauli-4 IC-POVM is denoted as $\textbf{P}=\{P(\textbf{a}) \}_\textbf{a}$ with $P(\textbf{a})\geq 0$ and $\sum_{\textbf{a}} P(\textbf{a}) =1$.

\subsection{Bidirectional Recurrent Neural Networks}
Recurrent neural network can  process effectively time series data in natural language processing (NLP) problems \cite{inproceedings,google_article,2018State}, which is extended to a bidirectional model by Schuster and Paliwal \cite{Schuster2002Bidirectional}. It can be trained without the limitation of using information just up to a preset future frame. In our experiments, we use a Bidirectional Gated Recurrent Unit (BiGRU) neural network, a network model developed by combining BiRNN network model and a gated recurrent unit \cite{Cho2014Learning}. BiRNN has been proved to  capture effectively the relationship of contextual semantics in natural language processing. GRU is able to prevent the gradient disappearance and gradient explosion of RNN. Furthermore, GRU uses less network parameters and it is  easier to be trained than long short-term memory (LSTM)  network \cite{1997Long}.

\begin{figure}[ht]
\centering
\includegraphics[scale=0.32]{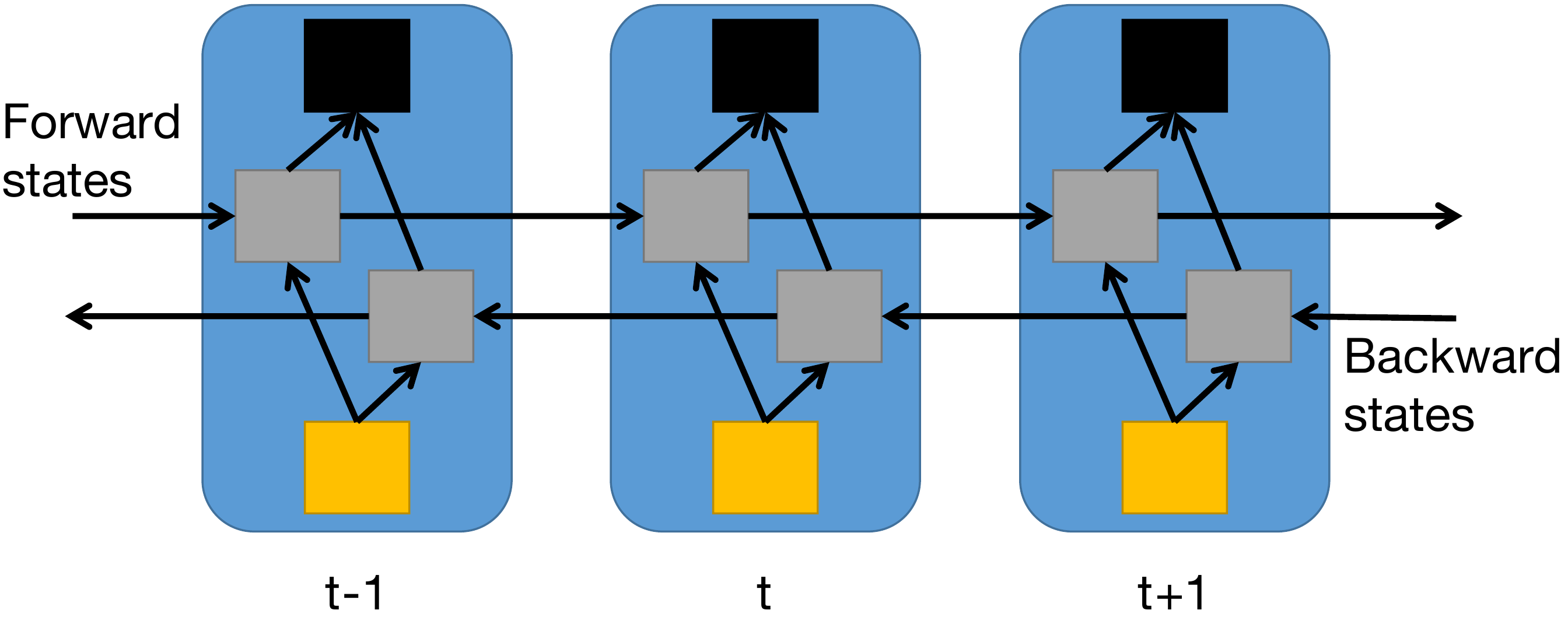}
\caption{\label{fig:fig1}The framework of general Bidirectional Recurrent Neural Networks shown unfolded in time for three time steps. The yellow blocks represent the inputs of quantum state samples and the gray blocks are the hidden states learned from forward and backward information flow of quantum state samples. The black blocks are the quantum state samples of next time step, which are fitted and predicted by the yellow and gray blocks.}
\end{figure}

In order to process QST by neural network with less quantum state measurement samples, we need to make better use of the limited training samples. According to the general bidirectional model, we divide the training samples into forward information flow and backward information flow. Given a batch of $N_s$ quantum state measurement samples $\textbf{E}= \{\textbf{a}_1,\textbf{a}_2,\textbf{a}_3,...,\textbf{a}_{N_s}\}$, we slice $\textbf{a}_i$ according to  time series $T$, $1\leq t \leq T$. The training procedure of this unfolded bidirectional network over time can be summarized as followings:
\begin{itemize}
\item[1)]\textbf{Forward flow training:} By receiving all input data for one time slice from $t=1$ to $T$, the hidden layer outputs $\mathop{h_t}\limits ^{\rightarrow} = f(a_t,\mathop{h_{t-1}}\limits ^{\rightarrow})$.
\item[2)]\textbf{Backward flow training:} By receiving time series data $a_i$ from $t=T$ to 1, the hidden layer outputs $\mathop{h_t}\limits ^{\leftarrow} = f(a_t,\mathop{h_{t+1}}\limits ^{\leftarrow})$.
\item[3)]\textbf{Consolidate information flow:} Outputs of the network training  through the joint bidirectional information flow $ P(a_{t+1})=f(\mathop{h_t}\limits ^{\rightarrow}, \mathop{h_t}\limits ^{\leftarrow})$.
\end{itemize}
This procedure is a general bidirectional model framework and the activation function $f$ has different forms in different recurrent unit. See Fig. \ref{fig:fig1} for the framework of general BiRNN for three time steps. The model is trained by minimizing the log-likelihood loss function $L$ over $\theta$, which is as following:
$$
L(\theta) = -\frac{1}{T} \sum_{t=1}^{T}log(P_\theta(a_{t})),
$$
where $\theta$ is a set of model parameters.

\subsection{Easy and hard quantum states}
Rocchetto  {\it et al.~}\cite{rocchetto2018learning} proposed to classify quantum states based on the hardness of sampling  probability distribution of  measurement result in a given basis. The Greenberger-Horne-Zeilinger (GHZ) quantum state has been discussed in many articles, which is a highly non-classical state, specified by $|\Psi_{GHZ}\rangle = \frac{1}{\sqrt{2}}(|0\rangle ^ {\otimes N} + |1\rangle ^ {\otimes N})$. As a simple quantum pure state, GHZ state is widely used in quantum communication protocols. This kind of states can be sampled easily in a large quantum system and we will learn and reconstruct it in this paper.

We consider a kind of random pure states as hard states which are generated by normalizing a $2^n$ dimensional complex vector drawn from  unit sphere according to Haar measure \cite{rocchetto2018learning}. This kind of hard states cannot be prepared easily on any realistic quantum computing device with a polynomial large circuit. It means that quantum measurement samples can only be obtained for few-qubit states. Up to now, the performances of reconstructing the probability distributions  \cite{rocchetto2018learning}, or the quantum state wavefunctions \cite{torlai2018neural} of these states are not very ideal.

\begin{figure*}[htbp]
\centering
\includegraphics[scale=0.45]{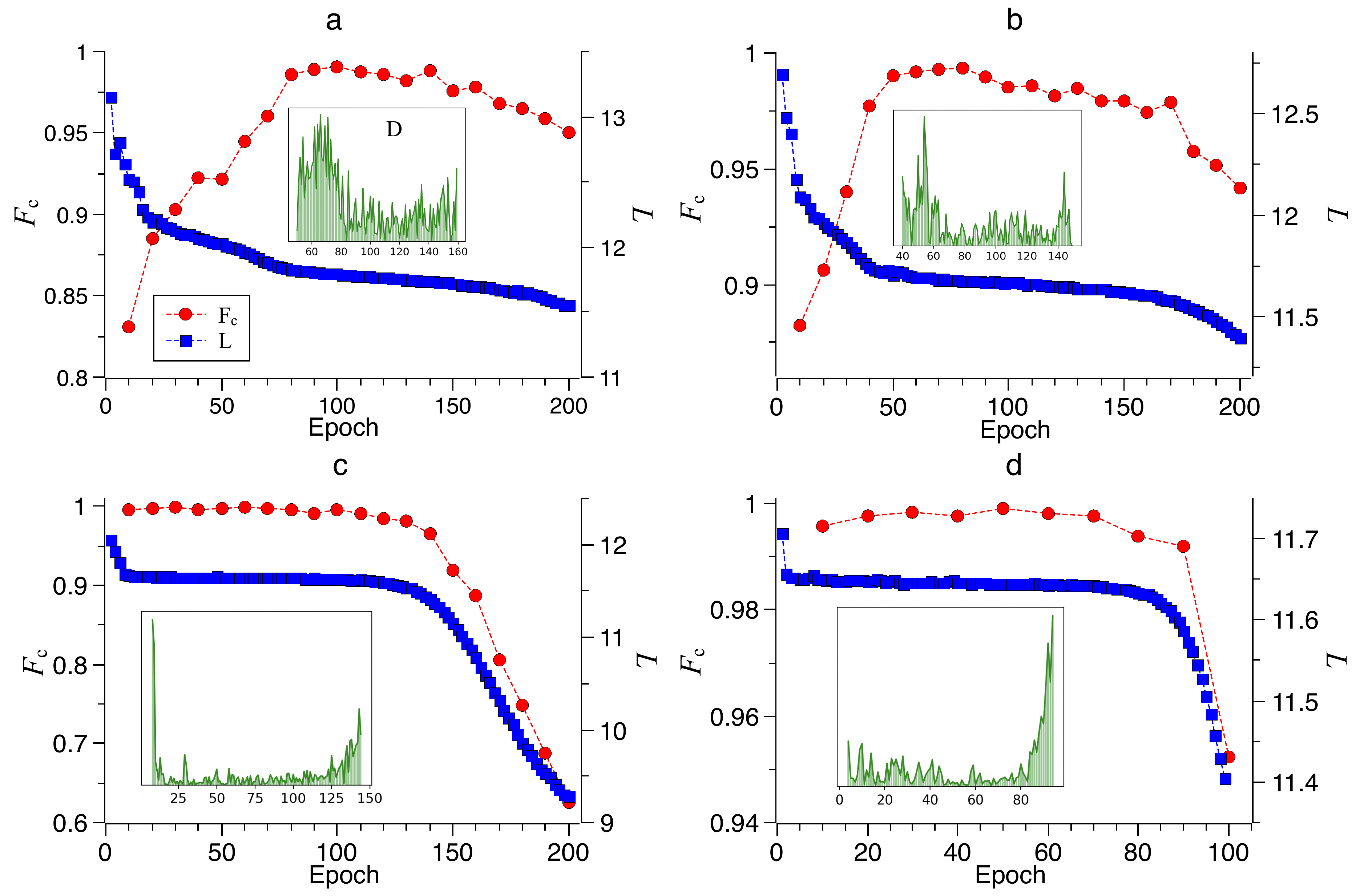}
\caption{\label{fig:fig2}The BiGRU generative model reconstructs 10-qubit GHZ quantum states. Subplots a, b, c, d show the results that use 500, 1000, 5000, and 30000 training samples, respectively. Red line and blue line are classic fidelity $F_c$ and loss function $L$. Green one is the degree of fluctuation $D$. We can  observe clearly that the following points: \textbf{(1)} $F_c$ first increases, then drops in the training process. This phenomenon can be explained by the overfitting problem. \textbf{(2)} Throughout the whole training process, when $F_c$ reaches its maximum, the fluctuation $D$ of the loss function $L$ is relatively small, thus $D$ can be used to overcome the overfitting problem. \textbf{(3)} As the number of training samples increases, the fluctuation $D$ is able to judge the performance of the model more accurately.}
\end{figure*}

\subsection{Learning standards}
 Quantum fidelity $F$ is a comprehensive measure of quantum state reconstruction. Let $\varrho_{GHZ}$ and $\varrho_{BiGRU}$ be two $N$-qubit quantum states, the quantum fidelity between them is defined as $$F(\varrho_{GHZ},\varrho_{BiGRU}) = \mathrm{tr}(\sqrt{\sqrt{\varrho_{GHZ}}\varrho_{BiGRU}\sqrt{\varrho_{GHZ}}}).$$ We say that $\varrho_{BiGRU}$ is a good representation of $\varrho_{GHZ}$ if the quantum fidelity $F \geq 1-\epsilon$ for a small $\epsilon > 0$. However, due to  exponential large dimension of the representation of density matrices, we can only calculate the quantum fidelity for reconstruction of small quantum systems. For large many-body quantum systems, we use  classical fidelity $F_c$ to evaluate the reconstruction performance, which is defined as $$F_c(\textbf P_{GHZ},\textbf P_{BiGRU})=\mathbb{E}_{\textbf a\sim \textbf P_{BiGRU}}\sqrt{\frac {P_{GHZ}(\textbf a)}{P_{BiGRU}(\textbf a)}}.$$
$F_c$ is a standard measure of proximity between two distributions. $P_{GHZ}(\textbf a)$, ${P_{BiGRU}(\textbf a)}$ are the exact probability and the probability generated by BiGRU that corresponding to the same quantum state measurement sample $\textbf a$,  respectively. Theoretically,   $F_c(\textbf{P}_{GHZ},\textbf{P}_{BiGRU})=1$ only if $\textbf{P}_{GHZ} = \textbf{P}_{BiGRU}$. In general,  $F_c$ will serve as a upper bound of $F$, meaning that a small error in $F_c$ will be amplified in $F$ \cite{Carrasquilla2019}.

The stop criterion is a difficult problem in training generative models. In this paper, we observe that a large number of training epoch which leads to  overfitting problem. Even if the loss function continues to decrease, the fidelity may decrease. Therefore, we cannot stop the training according to the loss function itself. Especially for the tomography of an unknown quantum state, how to use the fewest quantum state measurement samples and decide when to stop the network training to obtain the best fidelity will be the key issue. We propose a simple but effective metric for the stop criterion, i.e.,  degree of loss function's fluctuation: $D = \{d_i\}$ with $d_i= loss_i - loss_{i-1}$, which is the difference between the losses of two consecutive epochs. This metric can reflect the training quality of the network and help us to  find effectively the best quantum state reconstruction model with the fewest training samples. The detailed process is to select the range with the least fluctuation: $D=[d_j,...,d_{j+n}]$ and choose the trained network with the smallest loss in this range $BiGRU_{min(d_j,...,d_{j+n}})$, where $n$ is an appropriate value, e.g. 50.

\section{Numerical results}
Deep learning framework Pytorch is used in our numerical experiment.  We use a three-layer BiGRU in all experiments. The training is performed using backpropagation and the Adam optimizer \cite{Kingma2014Adam}, with initial learning rate of 0.001. The classical fidelity is calculated by 50000 generative samples drawn from the trained network.

We now turn our attention to the reconstruction of GHZ state. We show that the numerical results of reconstruction of 10-qubit GHZ quantum state using different number of training samples in Fig.~\ref{fig:fig2}. We find that no matter how large the number of training samples is, the training model will  overfit gradually as the number of training iterations increases. Even if we try to exceed the number of training samples mentioned by  Carrasquilla {\it et al.} in  \cite{Carrasquilla2019}, as shown in Fig.~\ref{fig:fig2}d, overfitting  occur also and result in poor reconstruction fidelity. In other words, how to prevent overfitting is the key point in this neural network QST technology. As far as we know, referencing the degree of loss function's fluctuation can assist us to find better models.

\begin{figure}[ht]
\centering
\includegraphics[scale=0.42]{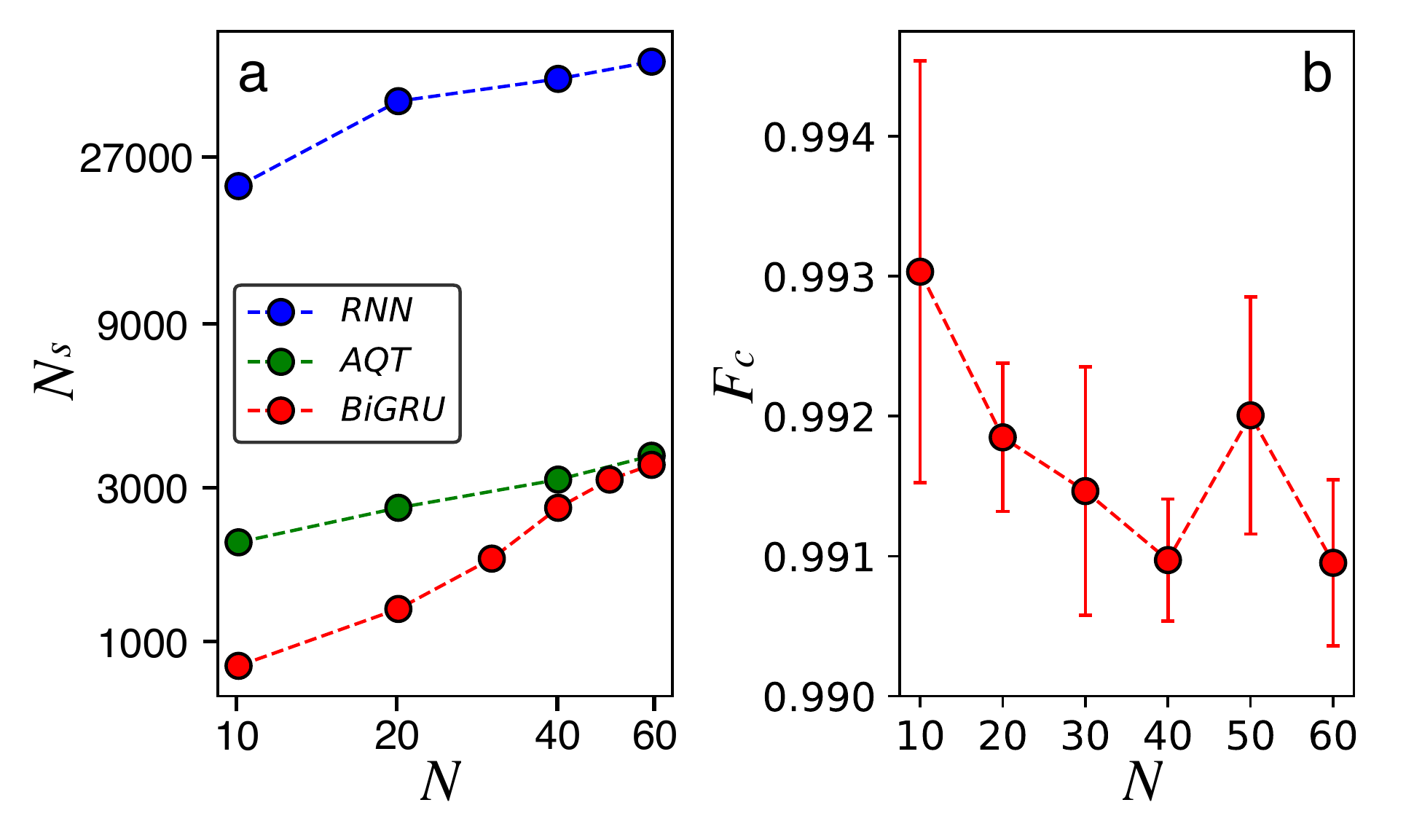}
\caption{\label{fig:fig3}The performance of reconstructing GHZ quantum states by BiGRU. Based on the  fluctuation $D$ to execute early stopping, we test BiGRU to reconstruct  GHZ quantum states with 10, 20, 30, 40, 50, 60 qubits. For each GHZ state, we obtain a network model and then sample five groups of generative samples to calculate five $F_c$ values. Each group contains 50000 generative samples. We require all five $F_c$ values achieve over 99\%. Subplot (a) : The log-log plot shows the necessary number of training samples ($N_s$ = 900, 1300, 1800, 2500, 3000, 3300), which is performed by BiGRU-QST. Compared with RNN-QST and AQT, our method needs the fewest training samples to reach over 99\% $F_c$.  Subplot (b): The plot shows the average $F_c$ over 5 calculations which is reconstructed by BiGRU-QST. The error bar stands for standard deviation.}
\end{figure}

In order to testify our method, we reconstruct GHZ states with system sizes ranging from $N = 10$ to 50 qubits.
The necessary number of training samples and the classical fidelities are shown in Fig.~\ref{fig:fig3}. Compared with the state-of-the-art RNN-QST \cite{Carrasquilla2019} and AQT \cite{Attentionbase2020}  methods, our BiGRU-QST method uses almost the fewest number of measurement samples to achieve over 99\% fidelity.  We can see also that a linear growth of the number of training samples with respect to the number of qubits.

\begin{figure}[ht]
\centering
\includegraphics[scale=0.6]{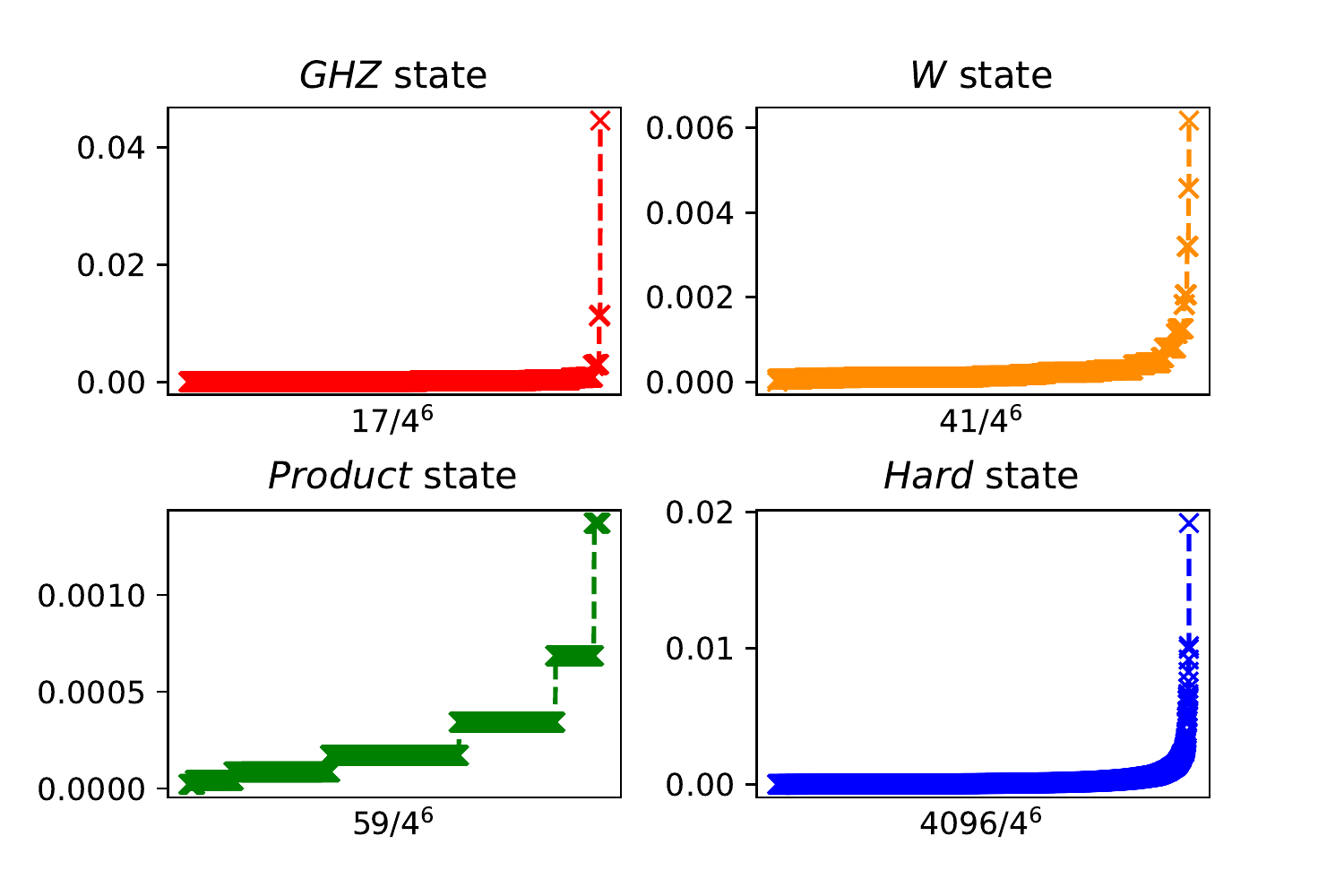}
\caption{\label{fig:fig4}
The distribution of measurement results of 6-qubit GHZ, W, Product and Hard states. The probability values shown have been sorted. There are 17, 41, 59 and 4096 different probability values for these states, respectively. More probability values will make the state more difficult to be reconstructed.}
\end{figure}

\begin{figure}[ht]
\centering
\includegraphics[scale=0.42]{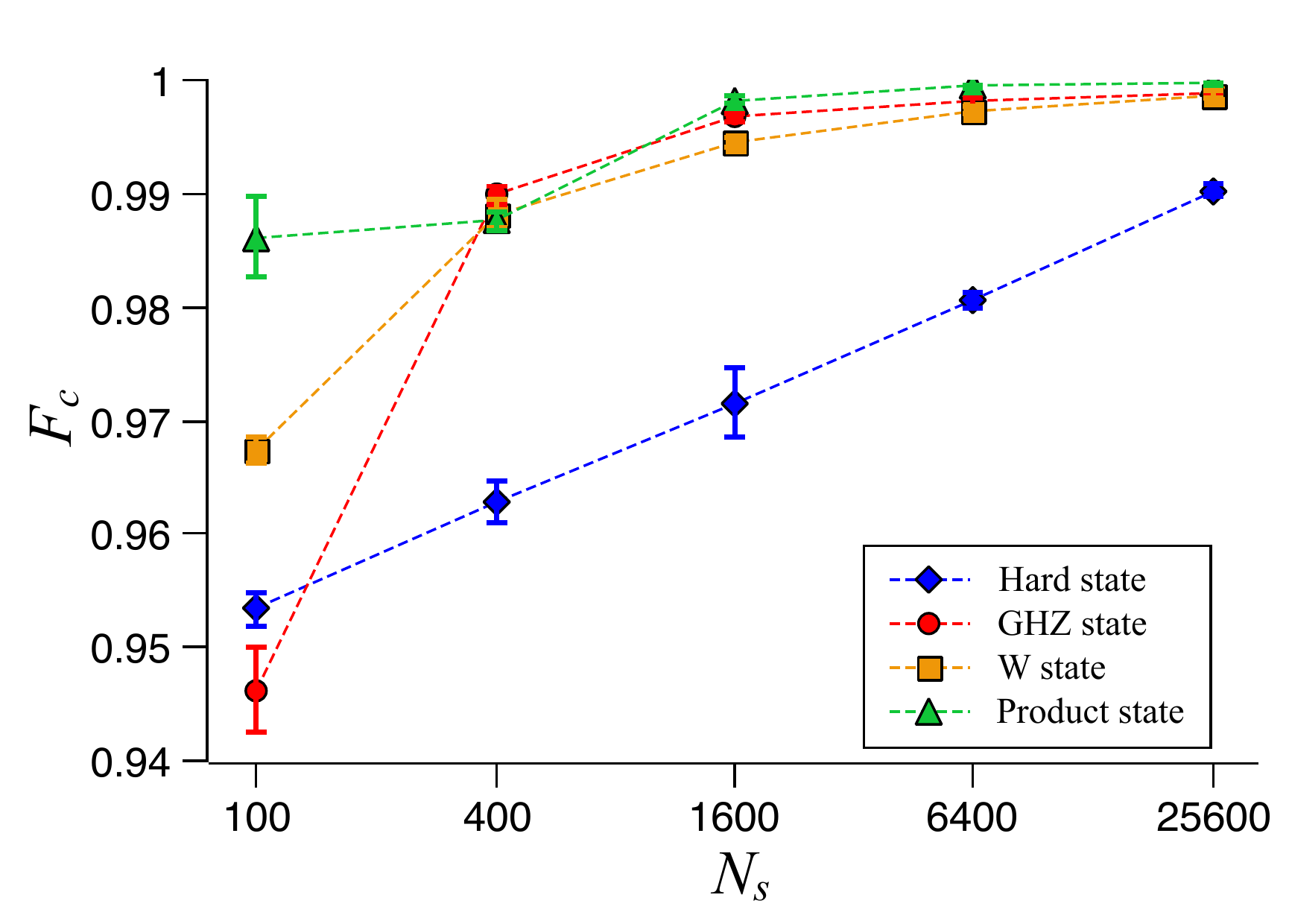}
\caption{\label{fig:fig5}Use BiGRU generative model to reconstruct 6-qubit Product state, hard state, GHZ state and W state with different numbers of training samples. For Product state, GHZ and W states, we use five groups of training samples to train five models and  sample 50000 generative samples from each models to calculate five $F_c$. For the hard state, we train five models using five groups of training samples produced by five different random pure states. The error bar stands for standard deviation. We find that the standard deviation will be smaller as the number of training samples increases.}
\end{figure}

When we use BiGRU to perform QST on the hard state, we find a limitation of this method. In our experiments, we prepare synthetic datasets mimicking experimental measurements of the 6-qubit exact distributions of quantum states $\textbf{P}_{GHZ}$, $\textbf{P}_{Hard}$, $\textbf{P}_W$ and $\textbf{P}_{product}$. The W state is written as $|\Psi_W\rangle = \frac{1}{\sqrt{N}}(|100...\rangle+|010...\rangle+...+|...001\rangle)$ and the Product state is a fully polarized state $|\Psi_{Product}\rangle = H^{\otimes}{^N} |0\rangle^{\otimes}{^N} = \frac{1}{\sqrt{2^N}}\sum_{i=0}^{2^N-1}|i\rangle$, where $|i\rangle$ is the computational basis. The probability distribution of measurement results of 6-qubit quantum system under the Pauli-4 measurement operators has  $4^6=4096$ probability values. The probability distributions after sorting for these four states are shown in Fig~\ref{fig:fig4}. Due to the relatively simple structure of W state, GHZ state and Product state, these $4^6$ probability values have only 41, 17 and 59 different values. In other words, these few probability values will occupy most of the quantum information and directly affect tomographic fidelity. In numerical results, we investigate $F_c$ as a function of $N_s$, the reconstructed results are shown in Fig.~\ref{fig:fig5}. We find that $F_c$ of Product state, W state and GHZ state  reach quickly 99\% fidelity with fewer measurement samples and  $F_c$ of the hard state increases much more slowly because the probability values of its measurement distribution are all different. In short word, we can get an approximate rule for QST method driven by these time series generative models, that is, the number of different probability values will seriously affect reconstruction results. The efficiency of QST will be better when the number of different measurement probability values is smaller.

\section{Conclusion}

In summary, the purpose of quantum state tomography is to reconstruct as complete quantum state as possible through limited quantum state measurement samples. Actually, it is impractical to perform QST in large quantum systems due to  exponential ``curse of dimensionality" inherent to  description of quantum states. Therefore, how to alleviate these scaling issues is what we are pursuing. The QST driven by neural networks helps us process specific quantum states to achieve high fidelity reconstructions with only a small amount of samples. In our work, we proposed a BiGRU neural network as generative model, which can effectively capture contextual associations in the field of natural language processing. In terms of QST by BiGRU, we encoded quantum state measurement samples into bidirectional information flow, and used BiGRU to capture the internal correlation of quantum information, and relied on early stopping skill to achieve better results compared with other neural network QST within the scope of our knowledge.

 We showed through numerical experiments that these time series generative models are capable of representing, especially Product state, GHZ state and W state, whose probability distributions have very few different values. BiGRU can use very few measurement samples to realize QST with high fidelity, over 99\%. On the other hand, the reconstruction of hard states with many different measurement probability values need more samples. Although performing hard state QST has no significant reduction of measurement samples, this method can be effectively used in QST, because the fidelity increases as the number of samples increases, which means this method is credible.

We have given a brief discussion and experimental results to show that these kinds of easy quantum states with few probability values can have a good QST performance by neural network method. In order to obtain more general conclusions, more experimental verification and theoretical explanation are needed.
In order to improve the versatility of this method in quantum state tomography technology, as further works,  we will investigate QST and find out relationship between the number of different probability values and the necessary number of measurement samples. Furthermore, we will also study characteristics of the probability distribution which may have a direct impact on neural network based tomography.

\begin{acknowledgments}
The authors are thankful to the anonymous referees and editor for their comments and suggestions
that greatly help to improve the quality of the manuscript. We thank Dr. Zhiming He for discussion to improve the presentations of some figures.
This work is supported by Guangdong Basic and Applied Basic Research Foundation (No. 2020A1515011204). S.Z. acknowledges supports in part from the National Natural Science
Foundation of China (Nos. 61602532).
\end{acknowledgments}

\bibliography{apssamp}

\providecommand{\noopsort}[1]{}\providecommand{\singleletter}[1]{#1}%
\begin{thebibliography}{29}
\expandafter\ifx\csname natexlab\endcsname\relax\def\natexlab#1{#1}\fi
\expandafter\ifx\csname bibnamefont\endcsname\relax
  \def\bibnamefont#1{#1}\fi
\expandafter\ifx\csname bibfnamefont\endcsname\relax
  \def\bibfnamefont#1{#1}\fi
\expandafter\ifx\csname citenamefont\endcsname\relax
  \def\citenamefont#1{#1}\fi
\expandafter\ifx\csname url\endcsname\relax
  \def\url#1{\texttt{#1}}\fi
\expandafter\ifx\csname urlprefix\endcsname\relax\def\urlprefix{URL }\fi
\providecommand{\bibinfo}[2]{#2}
\providecommand{\eprint}[2][]{\url{#2}}

\bibitem[{\citenamefont{Häffner et~al.}(2005)\citenamefont{Häffner, Hänsel,
  Roos, Benhelm, Chek-al kar, Chwalla, Körber, Rapol, Riebe, Schmidt
  et~al.}}]{Haffner}
\bibinfo{author}{\bibfnamefont{H.}~\bibnamefont{Häffner}},
  \bibinfo{author}{\bibfnamefont{W.}~\bibnamefont{Hänsel}},
  \bibinfo{author}{\bibfnamefont{C.~F.} \bibnamefont{Roos}},
  \bibinfo{author}{\bibfnamefont{J.}~\bibnamefont{Benhelm}},
  \bibinfo{author}{\bibfnamefont{D.}~\bibnamefont{Chek-al kar}},
  \bibinfo{author}{\bibfnamefont{M.}~\bibnamefont{Chwalla}},
  \bibinfo{author}{\bibfnamefont{T.}~\bibnamefont{Körber}},
  \bibinfo{author}{\bibfnamefont{U.~D.} \bibnamefont{Rapol}},
  \bibinfo{author}{\bibfnamefont{M.}~\bibnamefont{Riebe}},
  \bibinfo{author}{\bibfnamefont{P.~O.} \bibnamefont{Schmidt}},
  \bibnamefont{et~al.}, \bibinfo{journal}{Nature}
  \textbf{\bibinfo{volume}{438}}, \bibinfo{pages}{643} (\bibinfo{year}{2005}),
  \urlprefix\url{https://doi.org/10.1038/nature04279}.

\bibitem[{\citenamefont{Gross et~al.}(2010)\citenamefont{Gross, Liu, Flammia,
  Becker, and Eisert}}]{2010Quantum}
\bibinfo{author}{\bibfnamefont{D.}~\bibnamefont{Gross}},
  \bibinfo{author}{\bibfnamefont{Y.-K.} \bibnamefont{Liu}},
  \bibinfo{author}{\bibfnamefont{S.~T.} \bibnamefont{Flammia}},
  \bibinfo{author}{\bibfnamefont{S.}~\bibnamefont{Becker}}, \bibnamefont{and}
  \bibinfo{author}{\bibfnamefont{J.}~\bibnamefont{Eisert}},
  \bibinfo{journal}{Phys. Rev. Lett.} \textbf{\bibinfo{volume}{105}},
  \bibinfo{pages}{150401} (\bibinfo{year}{2010}),
  \urlprefix\url{https://link.aps.org/doi/10.1103/PhysRevLett.105.150401}.

\bibitem[{\citenamefont{Qi~Yin}(2018)}]{QiYin2018}
\bibinfo{author}{\bibfnamefont{C.-F. L. G.-C.~G.} \bibnamefont{Qi~Yin},
  \bibfnamefont{Guo-Yong~Xiang}}, \bibinfo{journal}{Chinese Physics Letters}
  \textbf{\bibinfo{volume}{35}}, \bibinfo{eid}{070302} (\bibinfo{year}{2018}),
  \urlprefix\url{http://cpl.iphy.ac.cn/EN/abstract/article_71189.shtml}.

\bibitem[{\citenamefont{T\'oth et~al.}(2010)\citenamefont{T\'oth, Wieczorek,
  Gross, Krischek, Schwemmer, and Weinfurter}}]{2010Permutationally}
\bibinfo{author}{\bibfnamefont{G.}~\bibnamefont{T\'oth}},
  \bibinfo{author}{\bibfnamefont{W.}~\bibnamefont{Wieczorek}},
  \bibinfo{author}{\bibfnamefont{D.}~\bibnamefont{Gross}},
  \bibinfo{author}{\bibfnamefont{R.}~\bibnamefont{Krischek}},
  \bibinfo{author}{\bibfnamefont{C.}~\bibnamefont{Schwemmer}},
  \bibnamefont{and}
  \bibinfo{author}{\bibfnamefont{H.}~\bibnamefont{Weinfurter}},
  \bibinfo{journal}{Phys. Rev. Lett.} \textbf{\bibinfo{volume}{105}},
  \bibinfo{pages}{250403} (\bibinfo{year}{2010}),
  \urlprefix\url{https://link.aps.org/doi/10.1103/PhysRevLett.105.250403}.

\bibitem[{\citenamefont{Moroder et~al.}(2012)\citenamefont{Moroder, Hyllus,
  T{\'{o}}th, Schwemmer, Niggebaum, Gaile, Gühne, and
  Weinfurter}}]{2012Permutationally}
\bibinfo{author}{\bibfnamefont{T.}~\bibnamefont{Moroder}},
  \bibinfo{author}{\bibfnamefont{P.}~\bibnamefont{Hyllus}},
  \bibinfo{author}{\bibfnamefont{G.}~\bibnamefont{T{\'{o}}th}},
  \bibinfo{author}{\bibfnamefont{C.}~\bibnamefont{Schwemmer}},
  \bibinfo{author}{\bibfnamefont{A.}~\bibnamefont{Niggebaum}},
  \bibinfo{author}{\bibfnamefont{S.}~\bibnamefont{Gaile}},
  \bibinfo{author}{\bibfnamefont{O.}~\bibnamefont{Gühne}}, \bibnamefont{and}
  \bibinfo{author}{\bibfnamefont{H.}~\bibnamefont{Weinfurter}},
  \bibinfo{journal}{New Journal of Physics} \textbf{\bibinfo{volume}{14}},
  \bibinfo{pages}{105001} (\bibinfo{year}{2012}),
  \urlprefix\url{https://doi.org/10.1088/1367-2630/14/10/105001}.

\bibitem[{\citenamefont{Baumgratz et~al.}(2013)\citenamefont{Baumgratz, Gross,
  Cramer, and Plenio}}]{2013Scalable}
\bibinfo{author}{\bibfnamefont{T.}~\bibnamefont{Baumgratz}},
  \bibinfo{author}{\bibfnamefont{D.}~\bibnamefont{Gross}},
  \bibinfo{author}{\bibfnamefont{M.}~\bibnamefont{Cramer}}, \bibnamefont{and}
  \bibinfo{author}{\bibfnamefont{M.~B.} \bibnamefont{Plenio}},
  \bibinfo{journal}{Phys. Rev. Lett.} \textbf{\bibinfo{volume}{111}},
  \bibinfo{pages}{020401} (\bibinfo{year}{2013}),
  \urlprefix\url{https://link.aps.org/doi/10.1103/PhysRevLett.111.020401}.

\bibitem[{\citenamefont{Biamonte et~al.}(2017)\citenamefont{Biamonte, Wittek,
  Pancotti, Rebentrost, Wiebe, and Lloyd}}]{biamonte2017quantum}
\bibinfo{author}{\bibfnamefont{J.}~\bibnamefont{Biamonte}},
  \bibinfo{author}{\bibfnamefont{P.}~\bibnamefont{Wittek}},
  \bibinfo{author}{\bibfnamefont{N.}~\bibnamefont{Pancotti}},
  \bibinfo{author}{\bibfnamefont{P.}~\bibnamefont{Rebentrost}},
  \bibinfo{author}{\bibfnamefont{N.}~\bibnamefont{Wiebe}}, \bibnamefont{and}
  \bibinfo{author}{\bibfnamefont{S.}~\bibnamefont{Lloyd}},
  \bibinfo{journal}{Nature} \textbf{\bibinfo{volume}{549}},
  \bibinfo{pages}{195} (\bibinfo{year}{2017}).

\bibitem[{\citenamefont{Situ et~al.}(2020)\citenamefont{Situ, He, Wang, Li, and
  Zheng}}]{situ2020quantum}
\bibinfo{author}{\bibfnamefont{H.}~\bibnamefont{Situ}},
  \bibinfo{author}{\bibfnamefont{Z.}~\bibnamefont{He}},
  \bibinfo{author}{\bibfnamefont{Y.}~\bibnamefont{Wang}},
  \bibinfo{author}{\bibfnamefont{L.}~\bibnamefont{Li}}, \bibnamefont{and}
  \bibinfo{author}{\bibfnamefont{S.}~\bibnamefont{Zheng}},
  \bibinfo{journal}{Information Sciences} \textbf{\bibinfo{volume}{538}},
  \bibinfo{pages}{193} (\bibinfo{year}{2020}).

\bibitem[{\citenamefont{He et~al.}(2021)\citenamefont{He, Li, Zheng, Li, and
  Situ}}]{he2021Variational}
\bibinfo{author}{\bibfnamefont{Z.}~\bibnamefont{He}},
  \bibinfo{author}{\bibfnamefont{L.}~\bibnamefont{Li}},
  \bibinfo{author}{\bibfnamefont{S.}~\bibnamefont{Zheng}},
  \bibinfo{author}{\bibfnamefont{Y.}~\bibnamefont{Li}}, \bibnamefont{and}
  \bibinfo{author}{\bibfnamefont{H.}~\bibnamefont{Situ}}, \bibinfo{journal}{New
  Journal of Physics, to appear, doi: 10.1088/1367-2630/abe0ae}
  (\bibinfo{year}{2021}).

\bibitem[{\citenamefont{Carleo and Troyer}(2017)}]{carleo2017solving}
\bibinfo{author}{\bibfnamefont{G.}~\bibnamefont{Carleo}} \bibnamefont{and}
  \bibinfo{author}{\bibfnamefont{M.}~\bibnamefont{Troyer}},
  \bibinfo{journal}{Science} \textbf{\bibinfo{volume}{355}},
  \bibinfo{pages}{602} (\bibinfo{year}{2017}).

\bibitem[{\citenamefont{Hartmann and Carleo}(2019)}]{hartmann2019neural}
\bibinfo{author}{\bibfnamefont{M.~J.} \bibnamefont{Hartmann}} \bibnamefont{and}
  \bibinfo{author}{\bibfnamefont{G.}~\bibnamefont{Carleo}},
  \bibinfo{journal}{Phys. Rev. Lett.} \textbf{\bibinfo{volume}{122}},
  \bibinfo{pages}{250502} (\bibinfo{year}{2019}),
  \urlprefix\url{https://link.aps.org/doi/10.1103/PhysRevLett.122.250502}.

\bibitem[{\citenamefont{Deng et~al.}(2017)\citenamefont{Deng, Li, and
  Das~Sarma}}]{deng2017quantum}
\bibinfo{author}{\bibfnamefont{D.-L.} \bibnamefont{Deng}},
  \bibinfo{author}{\bibfnamefont{X.}~\bibnamefont{Li}}, \bibnamefont{and}
  \bibinfo{author}{\bibfnamefont{S.}~\bibnamefont{Das~Sarma}},
  \bibinfo{journal}{Phys. Rev. X} \textbf{\bibinfo{volume}{7}},
  \bibinfo{pages}{021021} (\bibinfo{year}{2017}),
  \urlprefix\url{https://link.aps.org/doi/10.1103/PhysRevX.7.021021}.

\bibitem[{\citenamefont{Cai and Liu}(2018)}]{2018wang}
\bibinfo{author}{\bibfnamefont{Z.}~\bibnamefont{Cai}} \bibnamefont{and}
  \bibinfo{author}{\bibfnamefont{J.}~\bibnamefont{Liu}},
  \bibinfo{journal}{Phys. Rev. B} \textbf{\bibinfo{volume}{97}},
  \bibinfo{pages}{035116} (\bibinfo{year}{2018}),
  \urlprefix\url{https://link.aps.org/doi/10.1103/PhysRevB.97.035116}.

\bibitem[{\citenamefont{Fournier et~al.}(2020)\citenamefont{Fournier, Wang,
  Yazyev, and Wu}}]{leiPhysRevLett}
\bibinfo{author}{\bibfnamefont{R.}~\bibnamefont{Fournier}},
  \bibinfo{author}{\bibfnamefont{L.}~\bibnamefont{Wang}},
  \bibinfo{author}{\bibfnamefont{O.~V.} \bibnamefont{Yazyev}},
  \bibnamefont{and} \bibinfo{author}{\bibfnamefont{Q.}~\bibnamefont{Wu}},
  \bibinfo{journal}{Phys. Rev. Lett.} \textbf{\bibinfo{volume}{124}},
  \bibinfo{pages}{056401} (\bibinfo{year}{2020}),
  \urlprefix\url{https://link.aps.org/doi/10.1103/PhysRevLett.124.056401}.

\bibitem[{\citenamefont{Tang-Shi~Yao}(2019)}]{Tang-Shi}
\bibinfo{author}{\bibfnamefont{M.~Y. K.-J. Z. D.-Y. Y. C.-J. Y. Z.-L. F. H.-C.
  L. C.-H. L. L. W. L. W. Y.-G. S. Y.-J. S. H.~D.} \bibnamefont{Tang-Shi~Yao},
  \bibfnamefont{Cen-Yao~Tang}}, \bibinfo{journal}{Chinese Physics Letters}
  \textbf{\bibinfo{volume}{36}}, \bibinfo{eid}{068101} (\bibinfo{year}{2019}),
  \urlprefix\url{http://cpl.iphy.ac.cn/EN/abstract/article_105329.shtml}.

\bibitem[{\citenamefont{Torlai et~al.}(2018)\citenamefont{Torlai, Mazzola,
  Carrasquilla, Troyer, Melko, and Carleo}}]{torlai2018neural}
\bibinfo{author}{\bibfnamefont{G.}~\bibnamefont{Torlai}},
  \bibinfo{author}{\bibfnamefont{G.}~\bibnamefont{Mazzola}},
  \bibinfo{author}{\bibfnamefont{J.}~\bibnamefont{Carrasquilla}},
  \bibinfo{author}{\bibfnamefont{M.}~\bibnamefont{Troyer}},
  \bibinfo{author}{\bibfnamefont{R.}~\bibnamefont{Melko}}, \bibnamefont{and}
  \bibinfo{author}{\bibfnamefont{G.}~\bibnamefont{Carleo}},
  \bibinfo{journal}{Nature Physics} \textbf{\bibinfo{volume}{14}},
  \bibinfo{pages}{447} (\bibinfo{year}{2018}),
  \urlprefix\url{https://doi.org/10.1038/s41567-018-0048-5}.

\bibitem[{\citenamefont{Carrasquilla et~al.}(2019)\citenamefont{Carrasquilla,
  Torlai, Melko, and Aolita}}]{Carrasquilla2019}
\bibinfo{author}{\bibfnamefont{J.}~\bibnamefont{Carrasquilla}},
  \bibinfo{author}{\bibfnamefont{G.}~\bibnamefont{Torlai}},
  \bibinfo{author}{\bibfnamefont{R.~G.} \bibnamefont{Melko}}, \bibnamefont{and}
  \bibinfo{author}{\bibfnamefont{L.}~\bibnamefont{Aolita}},
  \bibinfo{journal}{Nature Machine Intelligence} \textbf{\bibinfo{volume}{1}},
  \bibinfo{pages}{200} (\bibinfo{year}{2019}),
  \urlprefix\url{https://doi.org/10.1038/s42256-019-0045-0}.

\bibitem[{\citenamefont{Vaswani et~al.}(2017)\citenamefont{Vaswani, Shazeer,
  Parmar, Uszkoreit, Jones, Gomez, Kaiser, and Polosukhin}}]{2017Attention}
\bibinfo{author}{\bibfnamefont{A.}~\bibnamefont{Vaswani}},
  \bibinfo{author}{\bibfnamefont{N.}~\bibnamefont{Shazeer}},
  \bibinfo{author}{\bibfnamefont{N.}~\bibnamefont{Parmar}},
  \bibinfo{author}{\bibfnamefont{J.}~\bibnamefont{Uszkoreit}},
  \bibinfo{author}{\bibfnamefont{L.}~\bibnamefont{Jones}},
  \bibinfo{author}{\bibfnamefont{A.~N.} \bibnamefont{Gomez}},
  \bibinfo{author}{\bibfnamefont{L.}~\bibnamefont{Kaiser}}, \bibnamefont{and}
  \bibinfo{author}{\bibfnamefont{I.}~\bibnamefont{Polosukhin}},
  \bibinfo{journal}{arXiv:1706.03762v5}  (\bibinfo{year}{2017}).

\bibitem[{\citenamefont{Cha et~al.}(2020)\citenamefont{Cha, Ginsparg, Wu,
  Carrasquilla, McMahon, and Kim}}]{Attentionbase2020}
\bibinfo{author}{\bibfnamefont{P.}~\bibnamefont{Cha}},
  \bibinfo{author}{\bibfnamefont{P.}~\bibnamefont{Ginsparg}},
  \bibinfo{author}{\bibfnamefont{F.}~\bibnamefont{Wu}},
  \bibinfo{author}{\bibfnamefont{J.}~\bibnamefont{Carrasquilla}},
  \bibinfo{author}{\bibfnamefont{P.~L.} \bibnamefont{McMahon}},
  \bibnamefont{and} \bibinfo{author}{\bibfnamefont{E.-A.} \bibnamefont{Kim}},
  \bibinfo{journal}{arXiv:2006.12469v1}  (\bibinfo{year}{2020}).

\bibitem[{\citenamefont{Ahmed et~al.}(2020)\citenamefont{Ahmed, Sanchez~Munoz,
  Nori, and Frisk~Kockum}}]{CGAN2020}
\bibinfo{author}{\bibfnamefont{S.}~\bibnamefont{Ahmed}},
  \bibinfo{author}{\bibfnamefont{C.}~\bibnamefont{Sanchez~Munoz}},
  \bibinfo{author}{\bibfnamefont{F.}~\bibnamefont{Nori}}, \bibnamefont{and}
  \bibinfo{author}{\bibfnamefont{A.}~\bibnamefont{Frisk~Kockum}},
  \bibinfo{journal}{arXiv:2008.03240v1}  (\bibinfo{year}{2020}).

\bibitem[{\citenamefont{Luchnikov et~al.}(2019)\citenamefont{Luchnikov, Ryzhov,
  Stas, Filippov, and Ouerdane}}]{luchnikov2019variational}
\bibinfo{author}{\bibfnamefont{I.~A.} \bibnamefont{Luchnikov}},
  \bibinfo{author}{\bibfnamefont{A.}~\bibnamefont{Ryzhov}},
  \bibinfo{author}{\bibfnamefont{P.-J.} \bibnamefont{Stas}},
  \bibinfo{author}{\bibfnamefont{S.~N.} \bibnamefont{Filippov}},
  \bibnamefont{and} \bibinfo{author}{\bibfnamefont{H.}~\bibnamefont{Ouerdane}},
  \bibinfo{journal}{Entropy} \textbf{\bibinfo{volume}{21}},
  \bibinfo{pages}{1091} (\bibinfo{year}{2019}), ISSN \bibinfo{issn}{1099-4300},
  \urlprefix\url{http://dx.doi.org/10.3390/e21111091}.

\bibitem[{\citenamefont{Rocchetto et~al.}(2018)\citenamefont{Rocchetto, Grant,
  Strelchuk, Carleo, and Severini}}]{rocchetto2018learning}
\bibinfo{author}{\bibfnamefont{A.}~\bibnamefont{Rocchetto}},
  \bibinfo{author}{\bibfnamefont{E.}~\bibnamefont{Grant}},
  \bibinfo{author}{\bibfnamefont{S.}~\bibnamefont{Strelchuk}},
  \bibinfo{author}{\bibfnamefont{G.}~\bibnamefont{Carleo}}, \bibnamefont{and}
  \bibinfo{author}{\bibfnamefont{S.}~\bibnamefont{Severini}},
  \bibinfo{journal}{npj Quantum Information} \textbf{\bibinfo{volume}{4}},
  \bibinfo{pages}{28} (\bibinfo{year}{2018}),
  \urlprefix\url{https://doi.org/10.1038/s41534-018-0077-z}.

\bibitem[{\citenamefont{Sutskever et~al.}(2014)\citenamefont{Sutskever,
  Vinyals, and Le}}]{inproceedings}
\bibinfo{author}{\bibfnamefont{I.}~\bibnamefont{Sutskever}},
  \bibinfo{author}{\bibfnamefont{O.}~\bibnamefont{Vinyals}}, \bibnamefont{and}
  \bibinfo{author}{\bibfnamefont{Q.~V.} \bibnamefont{Le}}, in
  \emph{\bibinfo{booktitle}{Proceedings of the 27th International Conference on
  Neural Information Processing Systems - Volume 2}} (\bibinfo{publisher}{MIT
  Press}, \bibinfo{address}{Cambridge, MA, USA}, \bibinfo{year}{2014}),
  NIPS'14, p. \bibinfo{pages}{3104–3112}.

\bibitem[{\citenamefont{Wu et~al.}(2016)\citenamefont{Wu, Schuster, Chen, Le,
  Macherey, Krikun, Cao, Gao, Macherey, Klingner et~al.}}]{google_article}
\bibinfo{author}{\bibfnamefont{Y.}~\bibnamefont{Wu}},
  \bibinfo{author}{\bibfnamefont{M.}~\bibnamefont{Schuster}},
  \bibinfo{author}{\bibfnamefont{Z.}~\bibnamefont{Chen}},
  \bibinfo{author}{\bibfnamefont{Q.}~\bibnamefont{Le}},
  \bibinfo{author}{\bibfnamefont{W.}~\bibnamefont{Macherey}},
  \bibinfo{author}{\bibfnamefont{M.}~\bibnamefont{Krikun}},
  \bibinfo{author}{\bibfnamefont{Y.}~\bibnamefont{Cao}},
  \bibinfo{author}{\bibfnamefont{Q.}~\bibnamefont{Gao}},
  \bibinfo{author}{\bibfnamefont{K.}~\bibnamefont{Macherey}},
  \bibinfo{author}{\bibfnamefont{J.}~\bibnamefont{Klingner}},
  \bibnamefont{et~al.}, \bibinfo{journal}{arXiv:1609.08144v2}
  (\bibinfo{year}{2016}).

\bibitem[{\citenamefont{{Chiu} et~al.}(2018)\citenamefont{{Chiu}, {Sainath},
  {Wu}, {Prabhavalkar}, {Nguyen}, {Chen}, {Kannan}, {Weiss}, {Rao}, {Gonina}
  et~al.}}]{2018State}
\bibinfo{author}{\bibfnamefont{C.}~\bibnamefont{{Chiu}}},
  \bibinfo{author}{\bibfnamefont{T.~N.} \bibnamefont{{Sainath}}},
  \bibinfo{author}{\bibfnamefont{Y.}~\bibnamefont{{Wu}}},
  \bibinfo{author}{\bibfnamefont{R.}~\bibnamefont{{Prabhavalkar}}},
  \bibinfo{author}{\bibfnamefont{P.}~\bibnamefont{{Nguyen}}},
  \bibinfo{author}{\bibfnamefont{Z.}~\bibnamefont{{Chen}}},
  \bibinfo{author}{\bibfnamefont{A.}~\bibnamefont{{Kannan}}},
  \bibinfo{author}{\bibfnamefont{R.~J.} \bibnamefont{{Weiss}}},
  \bibinfo{author}{\bibfnamefont{K.}~\bibnamefont{{Rao}}},
  \bibinfo{author}{\bibfnamefont{E.}~\bibnamefont{{Gonina}}},
  \bibnamefont{et~al.}, in \emph{\bibinfo{booktitle}{2018 IEEE International
  Conference on Acoustics, Speech and Signal Processing (ICASSP)}}
  (\bibinfo{year}{2018}), pp. \bibinfo{pages}{4774--4778}.

\bibitem[{\citenamefont{{Schuster} and
  {Paliwal}}(1997)}]{Schuster2002Bidirectional}
\bibinfo{author}{\bibfnamefont{M.}~\bibnamefont{{Schuster}}} \bibnamefont{and}
  \bibinfo{author}{\bibfnamefont{K.~K.} \bibnamefont{{Paliwal}}},
  \bibinfo{journal}{IEEE Transactions on Signal Processing}
  \textbf{\bibinfo{volume}{45}}, \bibinfo{pages}{2673} (\bibinfo{year}{1997}).

\bibitem[{\citenamefont{Cho et~al.}(2014)\citenamefont{Cho, Van~Merrienboer,
  Gulcehre, Bahdanau, Bougares, Schwenk, and Bengio}}]{Cho2014Learning}
\bibinfo{author}{\bibfnamefont{K.}~\bibnamefont{Cho}},
  \bibinfo{author}{\bibfnamefont{B.}~\bibnamefont{Van~Merrienboer}},
  \bibinfo{author}{\bibfnamefont{C.}~\bibnamefont{Gulcehre}},
  \bibinfo{author}{\bibfnamefont{D.}~\bibnamefont{Bahdanau}},
  \bibinfo{author}{\bibfnamefont{F.}~\bibnamefont{Bougares}},
  \bibinfo{author}{\bibfnamefont{H.}~\bibnamefont{Schwenk}}, \bibnamefont{and}
  \bibinfo{author}{\bibfnamefont{Y.}~\bibnamefont{Bengio}},
  \bibinfo{journal}{arXiv:1406.1078v3}  (\bibinfo{year}{2014}).

\bibitem[{\citenamefont{Hochreiter and Schmidhuber}(1997)}]{1997Long}
\bibinfo{author}{\bibfnamefont{S.}~\bibnamefont{Hochreiter}} \bibnamefont{and}
  \bibinfo{author}{\bibfnamefont{J.}~\bibnamefont{Schmidhuber}},
  \bibinfo{journal}{Neural Computation} \textbf{\bibinfo{volume}{9}},
  \bibinfo{pages}{1735} (\bibinfo{year}{1997}),
  \urlprefix\url{https://doi.org/10.1162/neco.1997.9.8.1735}.

\bibitem[{\citenamefont{Kingma and Ba}(2014)}]{Kingma2014Adam}
\bibinfo{author}{\bibfnamefont{D.}~\bibnamefont{Kingma}} \bibnamefont{and}
  \bibinfo{author}{\bibfnamefont{J.}~\bibnamefont{Ba}},
  \bibinfo{journal}{arXiv:1412.6980v9}  (\bibinfo{year}{2014}).

\end{thebibliography}

\end{document}